\documentclass[prl,twocolumn,showpacs,superscriptaddress,floatfix]{revtex4}
\usepackage{graphicx}
\usepackage{amssymb}
\usepackage{citesort}
\begin{document}
\title{ Hysteretic behavior at the collapse of the metal-insulator transition in BaVS$_3$ }
%
\author{N. Bari\v si\'c}
\affiliation{Institut de Physique de la mati\`{e}re complexe,
EPFL, CH-1015 Lausanne, Switzerland}
and
\affiliation{Institut of
Physics, Bijeni\v cka c. 46, Hr-10 000 Zagreb, Croatia}

\author{A. Akrap}
\affiliation{Institut de Physique de la mati\`{e}re complexe,
EPFL, CH-1015 Lausanne, Switzerland}

\author{H. Berger}
\affiliation{Institut de Physique de la mati\`{e}re complexe,
EPFL, CH-1015 Lausanne, Switzerland}

\author{L. Forr\'o}
\affiliation{Institut de Physique de la mati\`{e}re complexe,
EPFL, CH-1015 Lausanne, Switzerland}
\date{\today}
\begin{abstract}

Electrical resistivity as a function of temperature, pressure, and magnetic field was measured in high and low purity single crystals of BaVS$_3$ close to the critical pressure value $p_{cr}$$\approx$2 GPa, associated with the zero temperature insulator-to-metal (MI) transition. In the 1.8-2.0 GPa range, where the MI transition is below $\approx$20 K, one can observe a sudden collapse of the MI phase boundary upon increasing pressure and at fixed pressure a magnetic field induced insulator to metal transition. In high quality samples these features are accompanied by hysteresis in all measured physical quantities as a function of temperature and magnetic field. We ascribe these observations to the crossing of the MI and the magnetic phase boundaries upon increasing pressure.

\end{abstract}

\pacs{71.10.Hf, 71.30.+H, 72.80.Ga}\maketitle

One of the most intriguing problems of contemporary solid state physics is the behavior of materials in the proximity of a quantum critical point (QCP) where the second order phase transition reaches zero temperature by tuning one of the external parameters like pressure or magnetic field. This situation is often found for magnetic transitions in the metallic f-electron systems. It was recently pointed out that at high pressures BaVS$_3$, which is a strongly correlated 3$d^1$ system, shows a critical behavior \cite{Forro} in many respects analogous to that of the f-electron materials \cite{QCP}. In the case of BaVS$_3$, the broad band is one-dimensional and undergoes a Peierls transition, while the narrow band is dominant in the magnetic properties of the system. Close to the critical pressure $p_{cr}$, which drives the Peierls state to zero, the interplay of the two bands gives a very rich and intriguing physics. Recently, it has been shown that at somewhat lower than the critical pressure magnetic field suppresses the insulating Peierls state \cite{Fazekas}. The close vicinity of $p_{cr}$ is therefore investigated here in order to understand in more detail the underlying physics. We report hysteretic effects in high quality single crystals in the same pressure window both in temperature and magnetic field. This indicates that the low T metal to insulator (MI) transition is of weakly first order \cite{Thesis, QCP}.

BaVS$_3$ differs from the conventional d-electron systems, such as transition metal compounds, by its strongly anisotropic crystal structure built from VS$_3$ chains. In the high temperature metallic phase one vanadium electron is shared between three bands originating from a $d_z^2$ orbital pointing along the chain and doubly degenerate $e_g$ orbital tilted out of it. At $T_S$=230 K a Jahn-Teller transition occurs and the site degeneracy of the $e_g$ states is lifted. Below this temperature, the minimal model sufficient for interpretation of all known experimental results consists of two quarter filled bands \cite{QCP, Mitrovic} in which local coupling \cite{Lechermann} and commensurability effects \cite{QCP} play an important role \cite{Thesis}. The $d_z^2$ band is quasi 1D and broad in contrast to the narrow and more isotropic $e_g$ band \cite{Mitrovic, Band racuni, Prvi clanak}. Upon cooling, around 150 K in the metallic phase, 1d diffuse lines appear in X-ray scattering at a commensurate value \cite{Pouget}. Beyond the quasi one dimensionality (Q1D) these data show that the $d_z^2$ band shares the electrons with the $e_g$ band in a 1:1 ratio. The later is partially attributed to the strong onsite coupling \cite{Lechermann} which favors the equal distribution of electrons between the bands regardless of the overall number of electrons to share. This implies that the doped samples where the carrier density is altered, should also have 1:1 charge sharing which is not the case \cite{Fagot}. Thus the exact quarter filling of a stoichiometric BaVS$_3$ has to be partly attributed to commensurability effects \cite{Thesis, QCP}. In the metallic phase the magnetic susceptibility is Curie-like reflecting dominantly the paramagnetic contribution originating from the quasi localized narrow band ($e_g$) electrons.

At the metal-to-insulator transition $T_{MI}\approx$70 K the main order parameter is the lattice tetramerization along the chain associated with the condensation of the diffuse line into new supperlattice Bragg spots, and dimerizing the chains in the transverse directions \cite{Pouget, Inami}. In BaVS$_3$ this is accompanied by the cusp followed by a drop in the uniform magnetic susceptibility \cite{Mihaly, Matsuura}. This can be explained upon recalling that a commensurability of the order 4 \cite{Giamarchi} can enhance a spin density wave (SDW) in the $d_z^2$ band. Such a SDW couples directly to the $e_g$ spins by the above mentioned strong onsite coupling (Hund) \cite{QCP}. Indeed at $T_X\approx$30 K the spins order almost ferromagnetically along the chains and with the incommensurate period perpendicular to the chains \cite{Neutroni}. Since the distance between the chains is rather large (6,73 $\AA$) the coupling of the neighboring spins is weak. Thus the magnetic order below $T_X$ is presumably resulting from a compromise between the intrinsic ferromagnetism of the $e_g$ electrons and the SDW correlations of the $d_z^2$ electrons hereafter referred as avoided ferromagnetism (aFM).

In order to determine how does the rather intricate situation described above evolve in the vicinity of p $<$ $p_{cr}$, where the tetramerization is almost removed, resistivity was measured as a function of temperature, pressure and magnetic field in the experimental setup described elsewhere \cite {Fazekas}. Special care was given to distinguish between the high and low quality samples. Thus after obtaining crystals, by the previously established Tellurium flux method \cite {Kuriyaki}, a careful characterization was carried out. The high quality crystals satisfied the following criteria (i) resistivity measurements at ambient pressure were required to exhibit metallic behavior at high temperatures, a well defined change of the slope at $T_S$, a sharp MI transition and no sign of saturation of resistivity in the insulating phase, (ii) the resistivity measured at 2 GPa (corresponding to the metallic regime in the whole temperature range) needed to display a high residual resistivity ratio (RRR, ratio between resistivity at 300 K and 2 K) of around 50 (iii) magnetic susceptibility anisotropy was required to show clearly both low temperature transitions ($T_{MI}$ and $T_x$) with negligible a Curie tail at temperatures below 10 K. Both the high quality and the low quality (low RRR ~10) samples are taken into account in the discussion below. We suspect that the low quality character of the crystals originates from sulfur deficiency.

As in other Q1d materials, the MI transition is shifted by pressure to lower temperatures. The main reason is that pressure decreases the one-dimensionality of the system and reduces the nesting of the Fermi surface. In the pressure range of 1 bar to $\approx$1.75 GPa (hereafter referred as the "low pressure insulating region") the transition varies linearly with pressure as $T_{MI}\approx 70-29.15 \cdot$ p [GPa] and it would extrapolated to zero at 2.4 GPa. But instead, by increasing the pressure above 1.76 GPa ($T_{MI}$(p=1.76 GPa)$\approx$18.5 K, $T_{MI}$ gets close or smaller than the characteristic temperature $T_g\approx$15 K \cite{QCP}), the MI phase boundary collapses to zero within the narrow pressure range of 0.3 GPa around the critical pressure $p_{cr}$$\approx$2 GPa (Fig. 1). This indicates that the transition is more complex than a smooth pressure-induced 3 dimensionalization of the system.

\begin{figure}[tb]
\resizebox{3.2in}{!}{\includegraphics{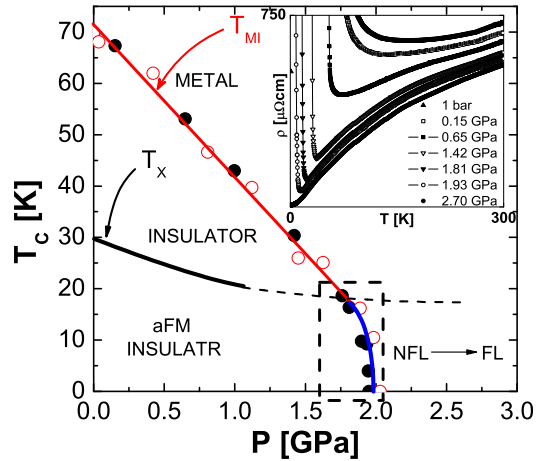}}
\caption {(Color online) The T-p phase diagram of BaVS$_3$ deduced from resistivity and magnetic measurements. The boxed window around $p_{cr}$ denotes the pressure range subject of this study. For the high purity crystals $T_{MI}$ marks the Peierls transition, but because of commensurability effect, the CDW has a SDW contribution, as well. $T_X$ denotes the incommensurate magnetic order (aFM) transverse to the chains. In low quality crystals $T_{MI}$ stands for the CDW transition (because of the lack of commensurability there is no SDW component) and $T_X$ for the ferromagnetic transition. Just above $p_{cr}$, where $T_{MI}$  goes to zero, both systems show a NFL behavior, which crosses over to FL \cite{QCP}.} \label{fig1}
\end{figure}

Such a conclusion is corroborated by the results obtained by applying magnetic field at pressures close below $p_{cr}$. In contrast to the low pressure insulating region, where the typical spin-Peierls magnetoresistance was recorded \cite{Kezsmarki}, in this case the magnetic field affects the system by \textquotedblleft switching off \textquotedblright the insulating phase independently of the sample quality. Conventionally this occurs through the coupling of the magnetic field to the spins of the $d_z^2$ electrons, \cite{Montambaux} which suppresses the CDW and/or SDW tetramerization. This was previously demonstrated on a low RRR sample \cite{Fazekas}. Here in Fig. 2 we report the data on a high RRR ($\approx$60) crystal. At 1.8 and 1.9 GPa one can observe a doubling of the transition. Both shift down slightly in temperature with increasing magnetic field. At 1.95 GPa ($T_{MI}=$ 7 K) a magnetic field of 12 T restores completely the metallic phase. In addition to the suppression of the MI phase transition with magnetic field, hysteresis is observed both for zero and finite magnetic field temperature cycles for all 3 pressures. The hysteresis is more pronounced for the phase transitions which are lower in temperature. It is important to note that besides the weak decrease of $T_{MI}$, the MI transition becomes ill-defined in magnetic field (the logarithmic derivative smears out) and collapses \cite{Thesis, Fazekas}.

\begin{figure}[tb]
\resizebox{3.2in}{!}{\includegraphics{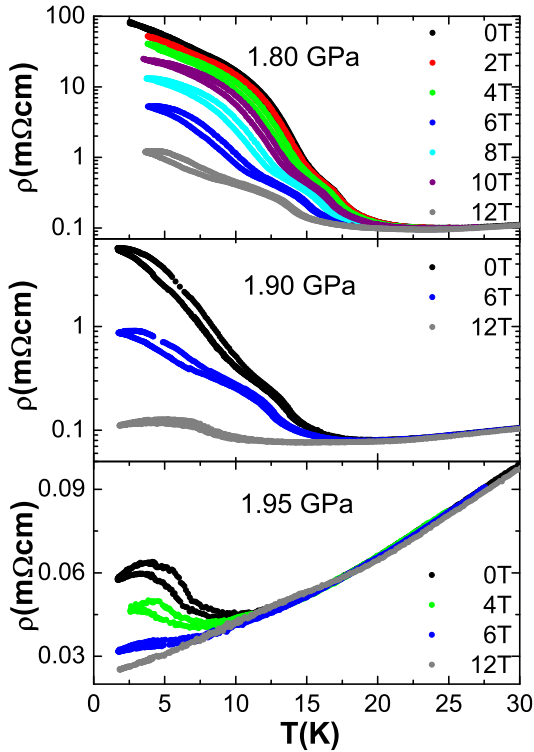}}
\caption{(Color online) Temperature cycles of the resistivity in magnetic field at 3 pressures close to $p_{cr}$. The doubling of the phase transition is related to the close proximity of $T_{MI}$ and $T_X$  at these pressures. Magnetic field suppresses $T_{MI}$  and weakly shifts it in temperature.} \label{fig2}
\end{figure}

It should be emphasized that the collapse of the MI phase boundary by pressure and the removal of the insulating phase by the magnetic field are common to both low \cite{Fazekas} and high quality BaVS$_3$  samples discussed here. Even more, in both cases the resistivity well above $p_{cr}$ up to $\approx$15 K is described by the same $T^n$ power-law, \cite {QCP} where n has a similar crossover from its non-Fermi liquid (NFL) value of n$\eqslantless$1.5 towards the Fermi Liquid (FL) value of n=2 \cite {QCP}. The substantial difference is that in the high quality samples hysteresis appears in resistance (Fig. 2), thermoelectric power \cite {Thesis} and magnetoresistance (Fig. 3). Since the appearance of hysteresis is the usual sign of a first order phase transition, it is reasonable to conclude for the high quality samples that in the pressure region where the $T_{MI}$$\approx$$T_X$ the order of MI transition is changed.

\begin{figure}[tb]
\resizebox{3.2in}{!}{\includegraphics{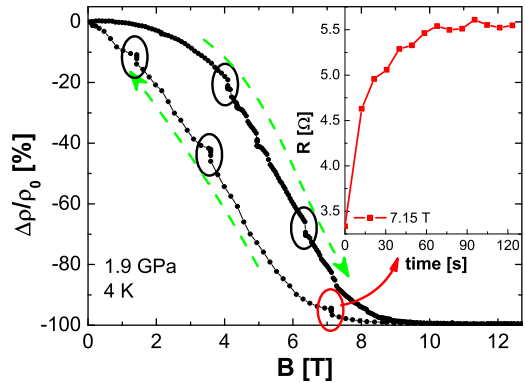}}
\caption{(Color online) Relative change of the resistivity at 1.9 GPa and at 4 K as a function of the magnetic field strength. The direction of the evolution of the magnetic field is indicated by green arrows. The field sweep was stopped several times at different field values as indicated by black and red circles. A slow relaxation of the resistivity with time was observed as shown in the inset at B=7.15 T.} \label{fig3}
\end{figure}

In order to understand better what happens in pure samples close to $p_{cr}$, the isothermal magnetoresistance is examined in more details just below $p_{cr}$. The result is shown in Fig. 3. The magnetoresistance is large and negative in agreement with Fig. 2. Several additional effects, related to the order of the transition, are observed in Fig. 3. First, the negative magnetoresistance is accompanied by a large hysteresis in magnetic field which closes at the critical field of $B_0$=9 T. Second, above $B_0$ the magnetoresistance is positive and hysteresis is absent, indicating that magnetic moments are ordered for large fields (B$>$B$_0$) and that only one phase is present. (Analogous behavior is observed in the metallic phase close above $p_{cr}$ \cite{QCP}). Third, if the sweep of the magnetic field is stopped the system relaxes on time scales of $\backsim$50 seconds. This is shown in the insets of Fig. 3, for the magnetic field of 7.15 T, on coming from higher fields. Worthy of note that even if the field sweep is stopped for extended periods, the upper and lower curves do not merge. Fourth, if the polarity of the magnetic field is changed the recorded curve is just a mirror image (in respect to the $\Delta$$\rho$/$\rho$ axis, not shown) of the curve plotted in Fig. 3. Such butterfly-like shapes are commonly observed in magnetoresistivity of the systems with ferromagnetic domains along the chains \cite{Leptir} due to their reordering in magnetic field. These measurements indicate strongly that the nature of the MI phase transition is of the first order and includes the kinetic aspects which are related to the slow movement of the domain walls.

\begin{figure}[tb]
\resizebox{3.2in}{!}{\includegraphics{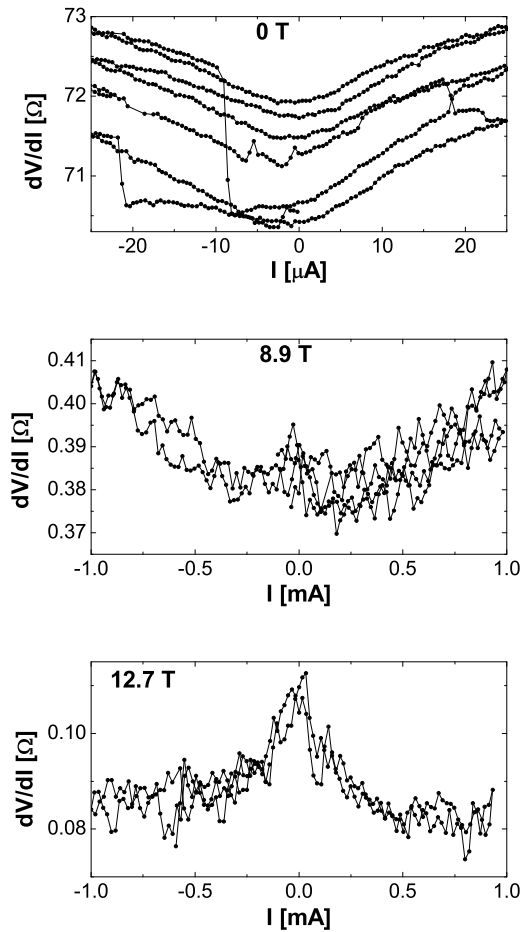}}
\caption{ The dV/dI characteristics, at T= 4 K, as a function of current for several magnetic fields, showing the instability of the system to the current.} \label{fig4}
\end{figure}

The coexistence of different phases in the region of hysteresis is supported by the I-V curves measured at 1.9 GPa and 4.2 K at different fixed magnetic fields (Fig. 4). At 0 T the dominant phase is the Peierls state, where the CDW order parameter is weak, and a Zener-type tunneling gives a V-shaped, nonlinear conductivity. With magnetic field one increases the number and size of magnetic domains and gradually a zero conductance peak develops characteristic for spin dependent transport (shown for 9 and 12.7 T). The increased noise level, and sudden switching to different conducting states with electric field is usual for the movement of domain walls \cite{Domains}. The slow change of the resistivity at fixed magnetic fields in the hysteresis loop is also the sign of the relaxation of the magnetic domains produced by sweeping the field.

All the observed features, e.g. the collapse of the MI transition under pressure, the field-induced insulator to metal transition, the appearance of hysteresis in high quality samples, the characteristic magnetic field $B_0$ etc., can be understood within the interplay of the Q1D $d_z^2$ and the more isotropic $e_g$ band. Since $T_{MI}$ is determined by the Peierls transition, it is expected that it decreases with pressure as the nesting conditions are weakened with increased interchain interactions. On the contrary $T_X$, associated with the magnetic transition of the $e_g$ band in the first place, should have only a weak pressure dependence. This latter prediction is supported by the measurements of H. Nakamura and T. Kobayashi \cite{unpublished} in a limited pressure range, which shows only a minute decrease of $T_X$. Several analogies performed on sister compounds point also in the direction that the magnetic ordering temperature varies more slowly with pressure than $T_{MI}$ \cite{QCP, Thesis}. For example, in Ba$_{1-x}$Sr$_x$VS$_3$ one can observe that chemical pressure shifts $T_X$ only slightly (up to a x$_{cr}$, where aFM changes to FM) \cite{Gauzzi}. Furthermore, studying the paramagnetic metal-ferromagnetic metal transition at $T_C$=42 K in BaVSe$_3$, it is learned that pressure does not change appreciably the value of $T_C$ \cite{Akrap, Thesis}. It is a reasonable scenario, that in the 1.8-2.0 GPa range the two phases cross while the interaction of the $e_g$ and $d_z^2$ electrons produces the doubling of the transition giving the first order character to the transition.

The corresponding interpretation of the hysteresis and the accompanying effects is the following. The fact that hysteresis was observed only in samples with high RRR indicates that it is due to the commensurability effects. Let us identify first the nature of the domains which occur in the B=0 situation. Obviously both kinds of domains are associated with comparable energies close to $p_{cr}$ (see Fig. 2) when $T_{MI}$ becomes of the order of $T_X$. In one kind of domain the commensurability energy prevails and the aFM order remains together with the commensurate CDW/SDW. In the other kind of domain the magnetic ordering of the $e_g$ electrons is ferromagnetic. The Weiss field corresponding to the magnetic condensation energy acts on $d_z^2$ electrons and splits their commensurate 2$k_F$ in two incommensurate ones. This suppresses the CDW/SDW similarly to the effect of the external field on the Peierls transition \cite{Montambaux}, i.e those domains are ferromagnetic and conducting. In other words the transition between the insulating phase and the conducting phase with ferromagnetism in presence of the commensurability effects is first order. Such competition of the $e_g$ and $d_z^2$ periodicities giving the first order character to the transition may be brought in a broad analogy to the CDW phase transitions with two periodicities which compete in presence of the commensurability \cite{Bjelis, Ayari}. Similar effects occur on polarizing the system by external magnetic field at a fixed pressure. In high RRR samples, the magnetic field forces the system towards the FM order, but with magnetic fields at our disposition it necessities also a relatively weak zero-field CDW/SDW amplitude. This range is typically 1.8-2.0 GPa. At a given pressure the magnetic field extends those domains at the expense of the $d_z^2$ CDW/SDW domains, in addition to polarizing the FM domains. It is thus reasonable to expect a hysteresis in resistivity as a function of T and B just as that observed in Figs. 2 and 3. Such interpretation is obviously independent of the sign of B. This is reflected in resistivity through the 2T broad butterfly-like hysteresis and the related relaxation phenomena.

In low RRR samples, presumably the sulfur deficiency changes $k_F$ from the commensurate to incommensurate value \cite{Fagot} causing the disappearance of the SDW component of the CDW and as a consequence, the magnetic order corresponds all along to a full FM at $T_X$. Consequently no hysteresis is observed in those samples although all other features, and in particular the (incommensurate) charge density wave (CDW) ordering, characteristic for BaVS$_3$, are still present.

In conclusion, we have demonstrated by magnetotransport measurements at high pressures that the M-I transition shows a first order character in the 1.8-2.0 GPa range. At these pressures the transition temperatures of the Q1d $d_z^2$ band with its CDW (and/or SDW) instability and the $e_g$ band with its FM order are close to each other. Due to the interplay of these competing orders, hysteresis is observed accompanied by the slow relaxation and noise of the magnetic domains. These features bring an additional richness to the physics of BaVS$_3$.

We acknowledge useful discussions with S. Bari\v si\'c, I.Kup\v ci\'c, T. Feher and A. Smontara. This work was supported by the Swiss and Croatian National Foundations for Scientific Research and theirs pools "Manep" and MoSES 035-0352826-2848 respectively.

\end{document}